\documentclass[twocolumn,prl,showpacs,floatfix,aps]{revtex4}
\usepackage{graphicx}
\usepackage{dcolumn}
\usepackage{bm}
\usepackage{color}

\def\ba{\begin{array}}
\def\ea{\end{array}}
\def\be{\begin{equation}\begin{array}{l}}
\def\ee{\end{array}\end{equation}}
\def\bea{\begin{equation}\begin{array}{l}}
\def\eea{\end{array}\end{equation}}
\def\f#1#2{\frac{\displaystyle #1}{\displaystyle #2}}

\begin{document}

\title{Tailoring  photon emission patterns in nanostructures}
\author{Shi-Fang Guo}\affiliation{Institute of Applied Physics and Computational
Mathematics, P. O. Box 8009, Beijing 100088, China}
\author{Su-Qing Duan}
\affiliation{Institute of Applied Physics and Computational
Mathematics, P. O. Box 8009, Beijing 100088, China}
\author{Yan Xie}
\affiliation{Institute of Applied Physics and Computational
Mathematics, P. O. Box 8009, Beijing 100088, China}
\author{Wei-Dong Chu}
\affiliation{Institute of Applied Physics and Computational
Mathematics, P. O. Box 8009, Beijing 100088, China}
\author{Wei Zhang {\footnote
{Author to whom any correspondence should be addressed, Email:
zhang$\_$wei@iapcm.ac.cn}
 }} \affiliation{Institute of Applied
Physics and Computational Mathematics, P. O. Box 8009, Beijing
100088, China}

\begin{abstract}
We investigate the photon emission in coupled quantum dots  based on
symmetry considerations. With the help of a new theorem we proved,
we reveal the origin of the various emission patterns, which is the
combinative symmetry in the time domain and spectrum domain. We are
able to tailor the emission patterns and obtain emission spectra
with odd harmonics only, even harmonics only, both odd and even
harmonic components, or even the quenching of all components. These
interesting emission patterns can be obtained in experiments by
careful design of the nanostructures, which are of many applications
in optical-electric nanodevices.
\end{abstract}

\maketitle

\section{\label{Sec:Intro}Introduction}
Photon emission in nanostructures plays a crucial role in modern
electronic/optical devices. Generating high-order harmonics is one
efficient up-conversion method for obtaining desired spectra (for
example, THz spectra) from sources with lower frequencies. During
the last several decades, much effort was made in the study of
generation of high-order harmonics in atomic/molecular systems and
nanostructures
\cite{1Ahn2007PRL,2Chassagneux2009Nature,3zhang2009PRB,agdot04,terzis,moiseyev2001},
as well as their applications. For instance, Ahn et al recently
proposed an approach for THz wave generation by high-order harmonic
wave based on semiconductor nanostructures driven by acoustic wave
\cite{1Ahn2007PRL}.  A scheme of electrically pumped
photonic-crystal THz laser \cite{2Chassagneux2009Nature} was
developed by Chassagneux et al. A method for THz wave generation by
Gigahertz wave was also suggested in \cite{3zhang2009PRB}.

Different harmonic components have interesting applications.  For
example, {\em even} harmonics were used as a test wave to diagnose
the fast time evolution of the current density
\cite{4Ferrante2005PLA}. Yet odd harmonics were often observed in
experiments in atomic and molecular systems. The appearance of odd
harmonics was attributed to the particular inversion symmetry in
central potential. There have been several studies on the dynamical
symmetry \cite{symmetry} and related selection rules \cite{select}
in high-order harmonics generation. People have tried to get even
harmonics by avoiding the selection rule \cite{5Xie2002PRA,
6Ferrante2004PLA, 8Pietro2007MO, 9Bavli1993PRA, 10Thomas2001PRL}.
Some studies focused on the symmetry breaking for obtaining the even
harmonics in molecules and atoms
\cite{9Bavli1993PRA,10Thomas2001PRL,8Pietro2007MO}. The symmetry
breaking may  be realized in the molecule with two nucleus of
different masses\cite{10Thomas2001PRL}.  The even harmonics were
found to appear in driven double quantum wells, where the potential
was not of inversion symmetry. Also the radiation may occur at
non-integer multiples of the fundamental frequency.

There have been many theoretical and experimental studies on
high-order harmonics, and most of them focused on the  atomic and
molecular systems \cite{11Wu2008OLA, 12Zhou2008PRA,
13Heslar2007IJQC, 14Le2007PRA,
15F2006PRL,16McPherson1987OSAB,17Huillier1991PhysB,18Protopapas1997RPP,19Salieres1999AAMOP,20Chang1997PRL,
21Schnurer1998PRL,22Salieres1998PRL,corso98}. In spite of many
studies on even harmonics generation, it appears that the deep
origin of different emission patterns is still unclear, and an
effective way of generating emission spectra of specific pattern is
lack. In this paper, we study the emission spectra of coupled
quantum dots (QDs), and find the symmetry origin of various emission
patterns, as well as methods for generating different emission
patterns including those with even harmonics. The main advantage of
QD (artificial atom) and coupled quantum dots (CQDs) (artificial
molecules) is their tunability. By carefully designing the material,
the growth process, or adding appropriate gate voltages etc., one is
able to adjust the energy levels/energy gaps in CQDs. One can
further design the structure of CQDs (the relative position of QDs,
the inter-dot distances, the hight and width of the tunneling
barriers) to tune the optical dipole between QDs. The optical
coupling between QDs can also be changed by tuning the polarization
of the incident light. Compared with one quantum dot with multiple
energy levels, the CQD system has the advantage of more tunability.
By making full use of the tunability of coupled nanostructures, we
propose an effective approach to tailor emission patterns. We reveal
the origin of appearance of odd/enen harmonics based on the symmetry
considerations. It turns out that the emission pattern is determined
not just by the inverse symmetry, but by a new type of symmetry,
which is the combinative symmetry in the time domain and spectrum
domain. Based on our findings, we are able to obtain emission
patterns with even harmonics only, odd harmonics only, or both even
and odd harmonics, and even disappearance of all harmonic
components. Our methods of generation of various emission patterns
in nanostructures have important applications.

\section{\label{Sec:HM} THEORETICAL FORMULISM}
 We consider a CQD system with one
energy level for each dot. The energy is $E_i$ for the state
$|i\rangle$ in the dot $i$, $i=1...N$. This CQD  is driven by an external 
field $E=F\cos(\omega_0 t) {\vec u} $, ${\vec u}$ unit vector.

Under the dipole approximation, our system is described by the
Hamiltonian
\begin{eqnarray}
H=\sum_i E_i |i\rangle \langle i |+\sum_{i\neq j} G_{ij} \cos
(\omega_0 t) |i \rangle \langle j| \label{H},
\end{eqnarray}
where $G_{ij}=F {\vec u}\cdot {\vec \mu_{ij}}$ are Rabi frequencies,
${\vec \mu_{ij}} =\langle i |e {\vec r} |j\rangle$ the dipole
between dot $i$ and dot $j$. The equation of motion for the density
matrix is written in the form
 \cite{23Narducci1990PRA}
\begin{eqnarray}
\frac{\partial \rho}{\partial
t}=-\frac{i}{\hbar}[H,\rho]-\Gamma\cdot\rho \label{D1},
\end{eqnarray}
where the last term describes possible dissipation effects (for
instance that from the spontaneous phonon  emission). We  set
$\hbar=1$ in the following. The time-dependent mean dipole moment
can be calculated as $P(t)=\langle er \rangle=\sum_{ij} \mu_{ij}
\rho_{ij}(t)$. With the help of Fourier transformation we can obtain
the emission spectrum $S(\omega)=|\int dt \exp(i\omega t)P(t)|^2$.
Now we present a theorem on the origin of the various emission
patterns. Several examples will be given.

{\bf Theorem }For a quantum system described by the Hamiltonian (1),
if there exists one symmetric operation Q, which is the time shift
$\theta: t \rightarrow t+T/2$ (or $-t+T/2$, $T=2\pi/\omega_0$)
combined with another operation $\Omega$ in spatial/spectrum domain,
i.e. $Q=\Omega\cdot \theta\ $, such that the initial condition and
the Hamiltonian  are invariant (or $H \rightarrow -H$), and the
dipole operator $\hat{P}$ has a definite parity, then the emission
spectrum contains no odd/even component if operator $\hat{P}$ is
even/odd.

The proof goes as following:

{\bf Proof } We first consider the case: under the operation Q, $H$
$\rightarrow$ $H$. The schrodinger equation $i\frac{d}{dt} |\psi
\rangle= H |\psi \rangle$ remains invariant under Q transformation.
If the initial condition remains unchanged (up to a phase), that is,
the initial condition $|\psi (t=0)\rangle$ satisfies
$|\widetilde{\psi} (t=0)\rangle\equiv Q |{\psi}
(t=0)\rangle=e^{-i\gamma}|{\psi} (t=0)\rangle$, then we have
$|\widetilde{\psi} (t)\rangle=e^{-i \gamma}|{\psi} (t)\rangle$.
Therefore
$P(t)=\langle\psi(t)|\hat{P}|\psi(t)\rangle=\langle\widetilde{\psi(t)}|\hat{P}|\widetilde{\psi(t)}\rangle
=\langle\psi(t+T/2)|\Omega^{-1}\hat{P}\Omega|\psi(t+T/2)\rangle=
\langle\psi(t+T/2)|\pm\hat{P}|\psi(t+T/2)\rangle=\pm P(t+T/2)$,
where "+" for even $\hat{P}$ and "-" for odd $\hat{P}$. Thus we have
$P(n\omega_0) \equiv \int dt e^{in\omega_0 t} P(t)=\pm \int dt e^{i
n\omega_0 (t-T/2)}P(t)=\pm (-1)^n P(n\omega_0)$.

Similarly for the case:  under the operation Q, $H$ $\rightarrow$
$-H$.
$P(t)=\langle\psi(t)|\hat{P}|\psi(t)\rangle=\langle\widetilde{\psi(t)}|\hat{P}|\widetilde{\psi(t)}\rangle
=\langle\psi(-t+T/2)|\Omega^{-1}\hat{P}\Omega|\psi(-t+T/2)\rangle=
\langle\psi(-t+T/2)|\pm\hat{P}|\psi(-t+T/2)\rangle=(\langle\psi(t-T/2)|)^*\pm\hat{P}(|\psi(t-T/2)\rangle)^*
=\langle\psi(t-T/2)|\pm\hat{P}|\psi(t-T/2)\rangle= \pm P(t-T/2)$,
where we have used the fact that $P(t)$ is real. Thus we also have
$P(n\omega_0)=\pm (-1)^n P(n\omega_0)$.

So we reach the results,

$P(n\omega_0)=0$ for odd n, if $\hat{P}$ is even (under $\Omega$);

$P(n\omega_0)=0$ for even n, if $\hat{P}$ is odd (under $\Omega$).

From above proof, one sees that the required initial condition is
$|\psi(t=0)\rangle=e^{i\gamma} Q |\psi(t=0)\rangle=e^{i\gamma}\Omega
|\psi(T/2)\rangle= e^{i\gamma}\Omega U(T/2)|\psi(0)\rangle$, where U
is the time evolution operator. So the initial state needs to be an
eigenvector of the operator $\Omega U(T/2)$. This initial condition
is not very convenient for practical use since the time evolution
operator is involved. Here we demonstrate that the initial condition
could be replaced with $|\psi(t=0)\rangle=e^{i\gamma}\Omega
|\psi(t=0)\rangle$, and the emission spectra have little change, as
long as $|E_i-E_j|/\omega_0$ is small.

Physically we can understand that for linear time evolution systems,
very small change of initial condition does not lead to big change
of the time evolution of the system. More precisely, from Floquet
theorem, we have $|\psi(t)\rangle=\sum_\alpha c_\alpha
e^{-i\varepsilon_\alpha t} |u_\alpha (t)\rangle$, with $|u_\alpha
(t)\rangle$ the time periodic Floquet state, $c_\alpha$ determined
by the initial condition. It is easy to see that
$P(\omega)=\sum_{\alpha,\alpha'}c^*_{\alpha'}c_{\alpha}K_{\alpha,\alpha'}(\omega)$,
where $K_{\alpha,\alpha'}(\omega)=\int dt e^{i\omega
t}e^{-i(\varepsilon_\alpha-\varepsilon_\alpha')t} u^*_{\alpha'}
(t)\hat{P} u_\alpha(t)$. So for two different initial conditions
with $c_\alpha,\widetilde{c}_\alpha$, such that
$|c_\alpha-\widetilde{c}_\alpha|<\eta$, $\eta$ a sufficient small
parameter, it is easy to show that
$|P(\omega)-\widetilde{P}(\omega)|<\sum_{\alpha,\alpha'}2\eta
K_{\alpha,\alpha'}$, here $P(\omega),\widetilde{P}(\omega)$ are the
emission spectra corresponding to the initial conditions with
$c_\alpha$, $c_\alpha'$ respectively.

Now we show that the difference between the initial conditions
satisfying $|\psi(t=0)\rangle=e^{i\gamma}\Omega |\psi(t=0)\rangle$
and $|\psi(t=0)\rangle=e^{i\gamma}Q |\psi(t=0)\rangle$ can be small
enough, if $|E_i-E_j|/\omega_0\ll 1$. The time evolution operator
satisfies the equation $i dU/dt=\hat{H} U$ with initial condition
$U(t=0)=I$, $I$ the identical operator. $\hat{H}=\hat{E}+G
\cos(\omega_0 t) \hat{\mu}$, where $\hat{E}=diag(E_1,...,E_N)$. The
symmetric matrix $\hat{\mu}$ can be diagonalized as
$A^{-1}\hat{\mu}A=diag(\lambda_1,...\lambda_N)$. Using the
transformation $U=AB\bar{U}A^{-1}$, $B=diag(e^{-i\lambda_1 G
f(t)},...,e^{-i\lambda_N G f(t)})$, $f(t)=\int_0^t d\tau
\cos(\omega_0 \tau)$, one sees that $id\bar{U}/dt=\bar{E}\bar{U}$,
where $\bar{E}=B^{-1}A^{-1}\hat{E}AB$, $\bar{U}(t=0)=I$. Then it is
clear that $|\bar{U}(T/2)-I|$ and
$|{U(T/2)}-I|=|A[\bar{U}(T/2)-I]A^{-1}|$ can be small enough, if
$|E_i-E_j|/\omega_0$ is small.  Since $Q= \Omega
U(T/2)=\Omega+\Omega U(T/2)-\Omega=\Omega+\Omega(U(T/2)-I)$,
standard perturbative calculation shows that the correction of
initial condition due to perturbation $\Omega(U(T/2)-I)$ is small
and one can use the initial condition
$|\psi(t=0)\rangle=e^{i\gamma}\Omega |\psi(t=0)\rangle$ for the
cases with $|E_i-E_j|/\omega_0\ll 1$. In the above proof the
dissipation hasn't been included, yet we would like to point out
that if the dissipation is small, the main feature of emission
spectra remains unchanged as verified by our numerical calculations
shown in the next section.

We have obtained one selection rule  in the perturbation regime (
$|E_i-E_j|/\omega_0\ll 1$) and it depends on the initial condition.
What may happen in the nonperturbation regime? In the regime with
large $|E_i-E_j|/\omega_0$ and $G/\omega_0$,  most quasienergies are
large and the corresponding quasienergy states in the expansion
$|\psi(t)\rangle=\sum_\alpha c_\alpha e^{-i\varepsilon_\alpha
t}|u_\alpha(t)\rangle$ lead to fast oscillation behavior, only the
quasienergy state with the smallest quasienergy
$\varepsilon_{\alpha_0}$ (modulo $\hbar \omega_0$) dominates, i.e.,
$|\psi(t)\rangle=c_{\alpha_0} e^{-i\varepsilon_{\alpha_0} t}
u_{\alpha_0}(t)$. If there exists one symmetric operation $Q$ such
that $QHQ^{-1}=H$, $Q i \frac{d}{dt} Q^{-1}=i \frac{d}{dt}$, then
the state $Q |u_{\alpha_0}\rangle$ is also a quasienergy state with
quasienergy $\varepsilon_{\alpha_0}$. If the quasienergy states are
nondegenerate, $Q |u_{\alpha_0}\rangle$ is just
$|u_{\alpha_0}\rangle$ (up to a phase). Straightforward calculation
shows that $P(t)=-P(t+T/2)$ (Note that $\hat{P}$ is odd in this
case). Therefore there is no even harmonics. One should notice that
this selection rule is insensitive to the initial condition.

To have a  clearer understanding of the physical picture, we study
the case of double dots (with two levels), since some explicit
solution could be obtained. The probability amplitudes
$\beta,\alpha$ for one electron in states $|1\rangle$ and
$|2\rangle$ satisfy the following equation
\bea
  \left( \ba{c} i \f{d\alpha}{dt}\\ i\f{d\beta}{dt}\ea \right)= \left( \ba{cc}
 \f{\Delta}{2} & G\cos(\omega_0 t) \\
G\cos(\omega_0 t) & -\f{\Delta}{2} \ea \right)  \left( \ba{c}
\alpha\\ \beta \ea \right) \, ,
 \,
 \eea
where $\Delta=E_2-E_1$. Define $A\equiv e^{iGf(t)}(\alpha+\beta)$
and $B\equiv e^{-iGf(t)}(\alpha-\beta)$. Then we have the equations
for $A$ and $B$ \bea i\f{dA}{dt}=\f{\Delta}{2} e^{i2Gf(t)}B \\
i\f{dB}{dt}=\f{\Delta}{2} e^{-i2Gf(t)}A. \eea The time-dependent
dipole P can be written as \be
P(t)=2\mu_{12}Re\{\alpha^*\beta\}=\mu_{12}Re\{A^*A-B^*B\}/2. \ee
After some algebric calculation, we have the equation for $P(t)$
\bea \f{dP}{dt}=\Delta Im\{(a^*+b^*)(a-b)e^{i2 G f(t)}\}\\ -\Delta^2
\int^t_0 d\tau \cos[2Gf(t)-2Gf(\tau)] P(\tau),\eea where $a,b$ are
the probability amplitudes for the electron in the two dots at
initial time $t=0$. We have made no approximation and the above
equation is exact. Here we see that the first term, i.e. the initial
condition is important for $\Delta\ll 1$. The dependence of emission
spectra on the initial states has also been noticed in
\cite{initial}. $P(t)$ becomes less sensitive to the initial
condition for large $\Delta$ as we've discussed in the
nonperturbation regime.   For a particular type of initial condition
$a=b$, i.e. $\rho_{11}(t=0)=\rho_{22}(t=0)=0.5$, the first term on
the right side of equation (6) vanishes. Then the equation can be
solved for small $\Delta$ and one finds that there is no odd
harmonic in the emission spectrum \cite{zhao96}. It agrees with our
theorem as also shown in Fig. (1b).

\section{\label{Sec:NU} Numerical results and discussions}
The above neat theorem is very powerful in the "harmonic
engineering". It leads to many important consequences and is very
useful for the applications in designing the optical emission
patterns of coupled nanostructures. In the following, we give a few
examples of the applications of our theory. In the numerical
calculations, the density matrix is obtained by solving equation
(\ref{D1}) through Runge-Kutta method with the time step of
$0.01/\omega_0$, total steps of 1500000 and appropriate initial
conditions. Emission spectra are obtained by numerically calculating
dipole through the evolution of density matrix elements
$\rho_{ij}(t)$. We set $\hbar \omega_0$ as the unit of energy.

\begin{figure}
\begin{center}
\includegraphics*[angle=0,height=0.43\textwidth,width= 0.4\textwidth,angle=-90]{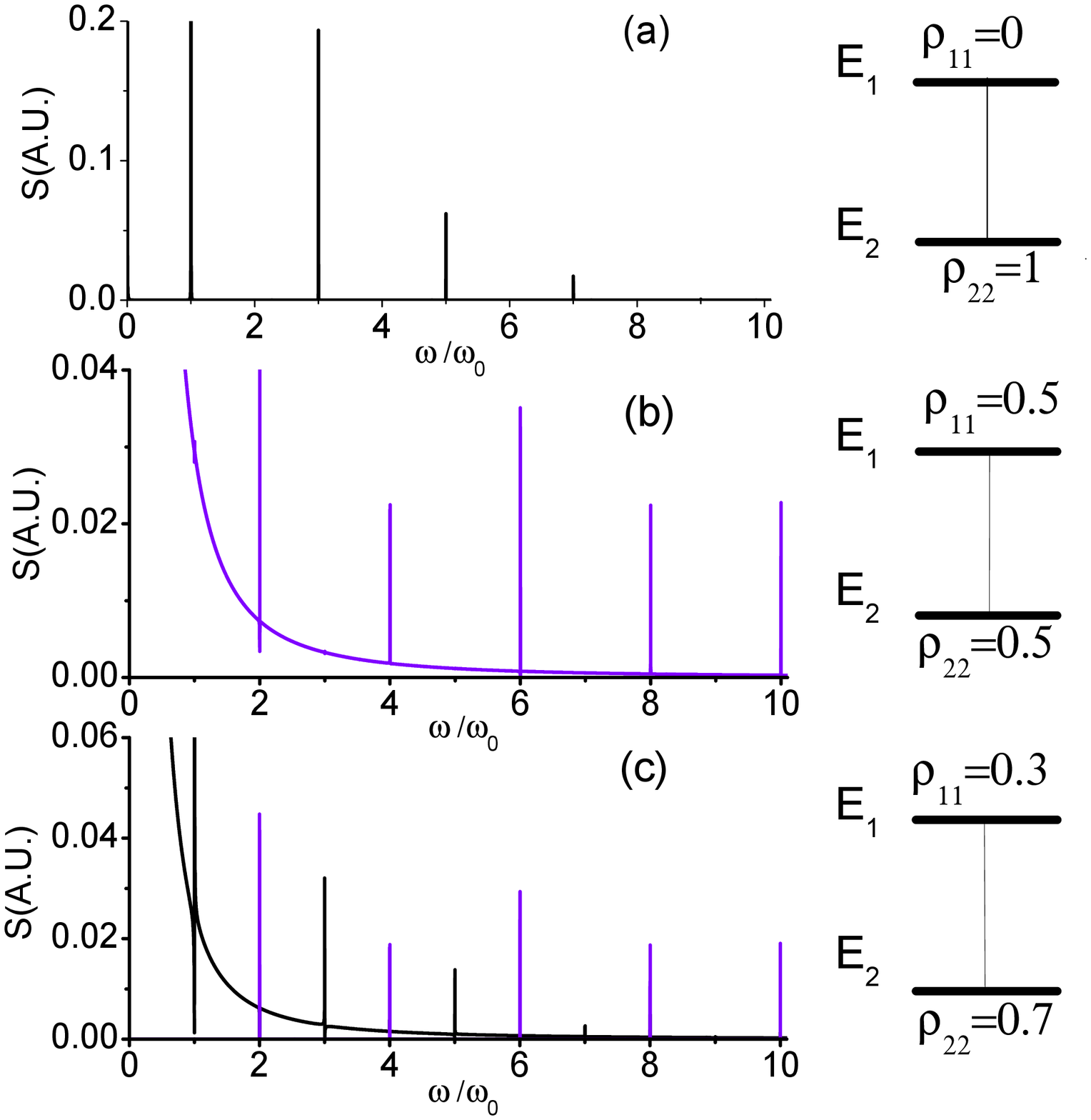}
\caption{Emission spectra in double dots with different initial
conditions. $\Delta=E_1-E_2=0.02$, $G_{12}=17.0$, and
$\Gamma=6.8\times 10^{-8}\omega_0$.} \label{FIG:CS}
\end{center}
\end{figure}

\begin{figure}
\begin{center}
\includegraphics*[angle=0,height=0.46\textwidth,width= 0.35\textwidth,angle=-90]{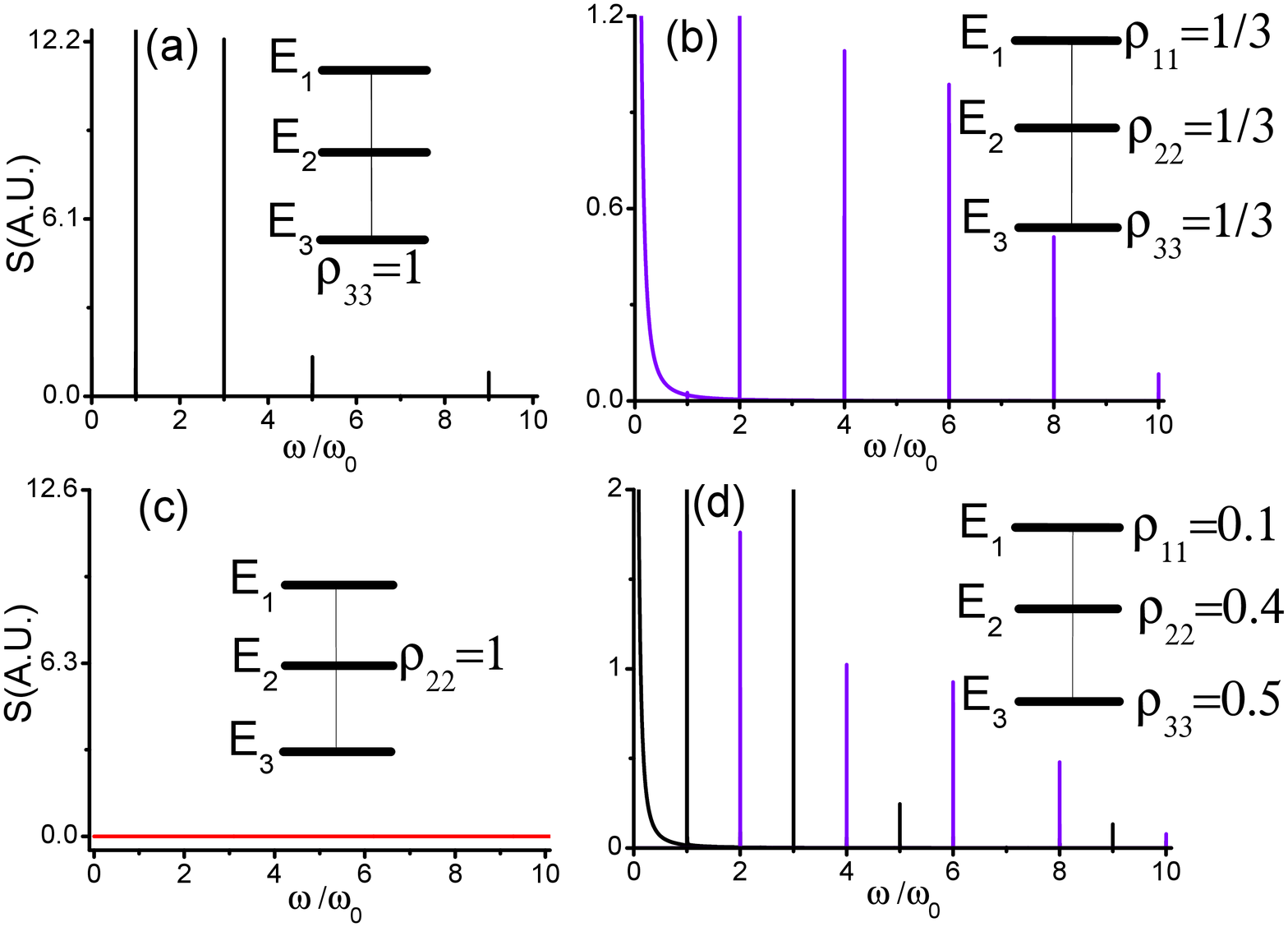}
\caption{Emission spectra in triple dots for different initial
conditions. The thin lines between the levels (in each dot) indicate
the optical couplings between the dots with value $G=15.25$.
$E_1-E_2=E_2-E_3=0.02$. $\Gamma=6.8\times 10^{-8}\omega_0$.}
\label{FIG:BE}
\end{center}
\end{figure}

\begin{figure}
\begin{center}
\includegraphics*[angle=0, height=0.36\textwidth,width= 0.45\textwidth]{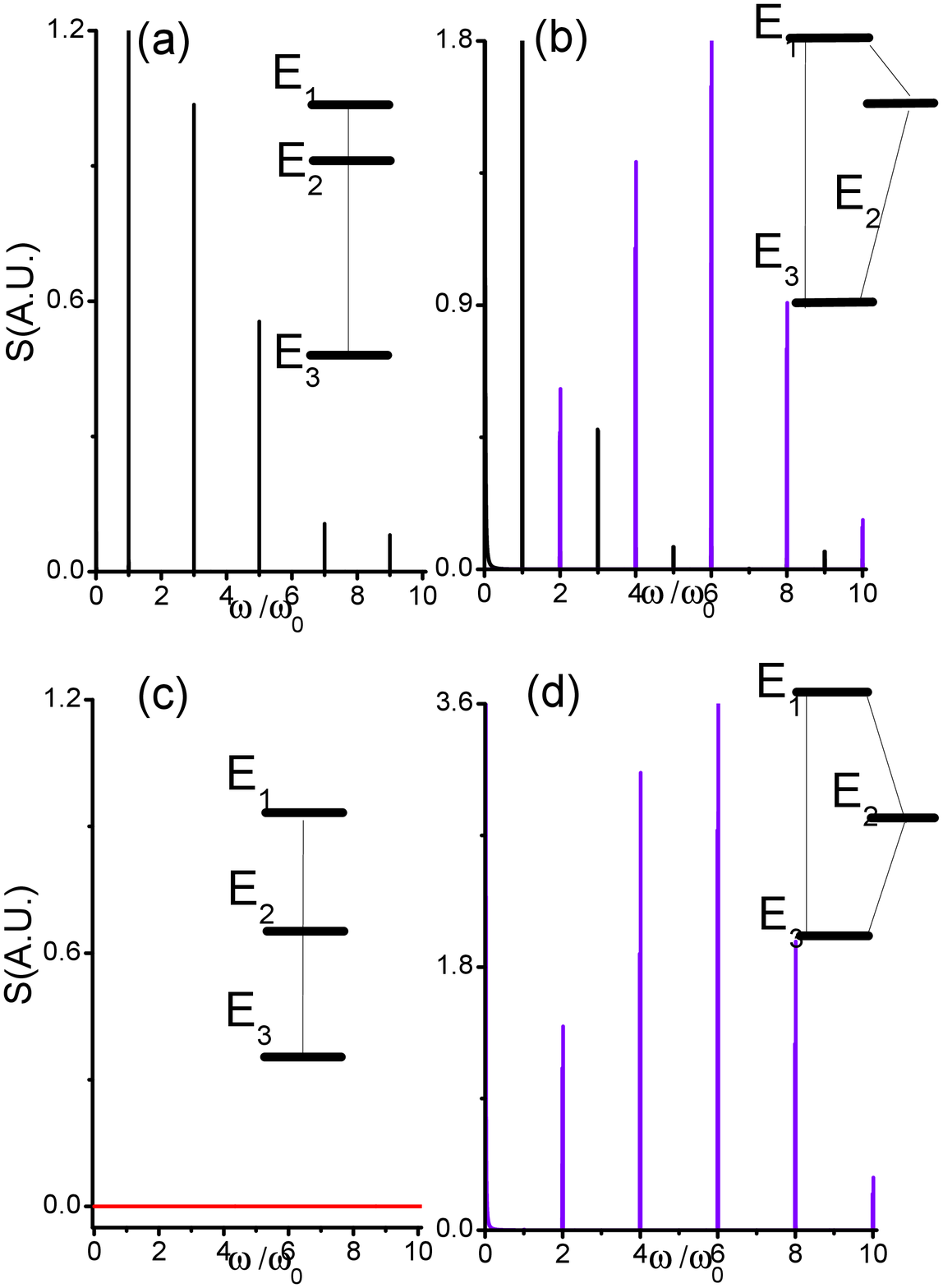}
\caption{ Emission spectra in triple dots with same initial
condition $\rho_{22}=1.0$, but with different structures. The thin
lines between the levels (in each dot) indicate the optical
couplings between the dots with value $G=5.0$. (a) and (c) are the
cases with chain structure. (b) and (d) are of loop structures. In
(a) and (b), $E_1-E_2=0.01$, $E_2-E_3=0.02$; in (c) and (d)
$E_1-E_2=E_2-E_3=0.02$. $\Gamma=6.8\times 10^{-8}\omega_0$.
}\label{FIG:3}
\end{center}
\end{figure}

\begin{figure}
\begin{center}
\includegraphics*[angle=0,height=0.48\textwidth,width= 0.35\textwidth,angle=-90]{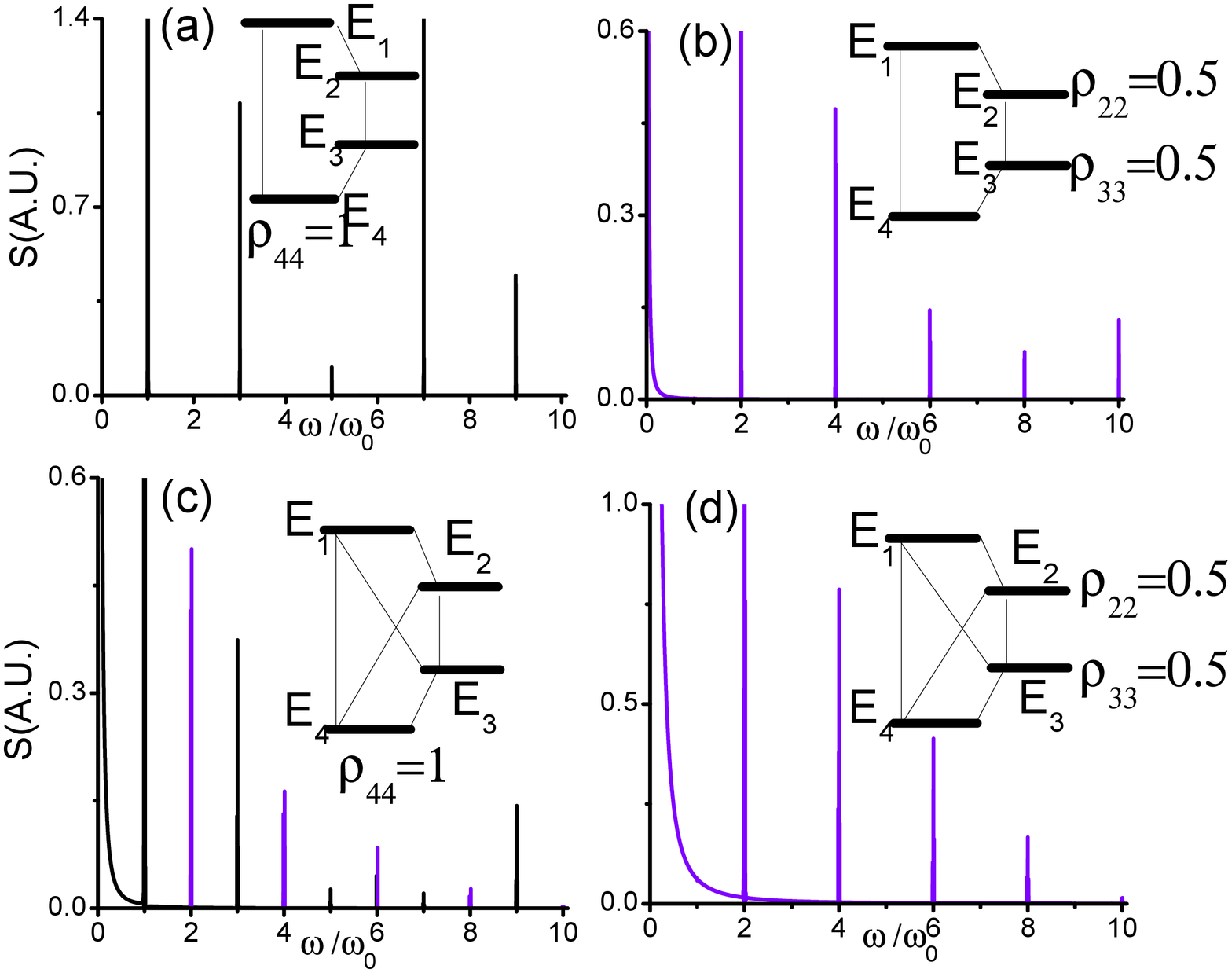}
\caption{Emission spectra in quadruple dots with different
structures/initial conditions. The thin lines between the levels (in
each dot) indicate the optical couplings between the dots with value
$G=6.3$. (a) and (b) are the cases with loop structure. (c) and (d)
are of additional cross couplings.  $E_1-E_2=E_3-E_4=0.008$,
$E_2-E_3=0.02$. $\Gamma=6.8\times 10^{-8}\omega_0$. }\label{FIG:4}
\end{center}
\end{figure}

{\bf Double dots} We first consider the system of double dots with
two levels. One can define Q as the time shift  $\theta: t
\rightarrow t+T/2$ combined with $\Omega_1: c_1\rightarrow -c_1$
(here and in the following $c_j, j=1,...,N$, refers to the
annihilation operator for state $|j\rangle$ in the dot j). It is
easy to see that under such transformation H is invariant,
$\hat{P}=G(c_1^+c_2+c_2^+c_1)$ is odd, and the initial condition
$\rho_{22}=1$ is invariant under transformation $\Omega_1$. So there
is no even component and odd peaks appear only, as seen in Fig. (1a)
and observed in many previous literatures.
 As our theory points out, the emission patterns
depend on the initial condition. We can apply our theory with
$\Omega_2: c_1\longleftrightarrow c_2$ for the initial condition
$\rho_{11}=\rho_{22}=0.5$. In this case $\hat{P}$ is even. Thus
there is no odd harmonic and only even harmonics appear as seen in
Fig. (1b). (Note that these peaks may appear as very small split
from exact even components.) It also agrees with our previous
analytical perturbation results. It is a very striking fact that
even the peak for $\omega_0$ (corresponding to Rayleigh peak, which
is often very pronounced in the usual emission spectra) disappears
due to the particular symmetry. As observed in Fig. (1c), there are
both odd and even components for the initial condition
$\rho_{11}=0.3,\rho_{22}=0.7$. This pattern is related to the fact
that both symmetries $\Omega_1 \cdot \theta$ and $\Omega_2 \cdot
\theta$ are broken.

{\bf Triple dots} We first show we can use different initial
conditions to realize quite different emission patterns. These
results are displayed in Fig. 2. It is seen that the emission
spectra may show odd only, even only, or both even and odd
components. In particular, the emission may be completely quenched
under appropriate condition. These results are based on our theorem.
The symmetry property related to $\Omega_1: c_2\rightarrow -c_2$
leads to the disappearance of even harmonics in Fig. (2a). Symmetry
$Q=\Omega_2\cdot\theta$, $\Omega_2: c_1\longleftrightarrow c_3$,
results in the spectrum without odd harmonics shown in Fig. 2(b).
Quite interestingly,  both $\Omega_1$ and $\Omega_2$ are satisfied
in Fig. (2c), which leads to the complete quenching of emission.
This interesting phenomenon is a direct consequence of symmetry and
is independent of the frequency of incident light. Thus it is
different from the quenching of emission due to coherent trapping.
The breaking of both symmetries $\Omega_1$ and $\Omega_2$ leads to
emission with odd and even components as shown in Fig. (2d).

Figure 3 shows the dependence of emission patterns on the structure
of nanosystems. Figs. (3a) and (3c) show systems with chain
structures and Figs. (3b) and (3d) show systems with loop
structures. Systems (c) and (d) are of symmetric energy levels,
which is absent in  systems (a) and (b). The different emission
patterns are the direct consequences of the symmetries generated by
$\Omega_1$ for (a), $\Omega_2$ for (d), $\Omega_1$ and $\Omega_2$
for (c) and none of $\Omega_1$ or $\Omega_2$ for (b).

{\bf Quadruple dots} Our theorem is very powerful and can be applied
to more general/complex nanostructures. Here we give one example of
the tailoring the emission patters of  quadruple dots. Careful
design of structures of CQDs, or initial conditions leads to
interesting emission spectra as shown in Fig. 4. The generators of
the related symmetries are: $\Omega_1: c_2\rightarrow -c_2,
c_4\rightarrow -c_4$ for (4a); $\Omega_2: c_2\longleftrightarrow
c_3, c_1\longleftrightarrow c_4$ for (4b) and (4d); None for (4c).

Our theorem on the essential roles of symmetry for the emission
patterns emphasizes the dynamic symmetry of the system, instead of
spatial symmetry. For a system with spatial inversion symmetry
$\Omega_0$, there may be odd harmonics only since the Hamiltonian is
invariant under the symmetric operation  $Q=\Omega_0\cdot\theta$ and
$\hat{P}$ is odd under $\Omega_0$. It is clear that dynamic symmetry
is more basic. Our theory points out that the emission spectra
depend not only on Floquet states (in particular their parity), but
also on the initial state. Moreover we've found nanostructures  with
more general symmetries, which lead to more interesting emission
patterns. Our studies suggest the effective methods for generation
the emission spectra with odd harmonics only, even harmonics only,
both odd and even harmonics, or even quenching of all components.
While previous work studied only the generation of even harmonics in
addition to the odd harmonics (i.e. both odd and even harmonics
appear). Our theory can be applied to more complex configurations.

We have considered the high-order harmonics generation from symmetry
point of view. In experiments, the symmetric configuration could be
realized by careful design the nanostructures. For instance, the
energy levels can be adjusted by changing the confining potential.
They could also be tuned by applying appropriate gate voltages. The
optical couplings among the dots could be tuned by changing the
inter-dot barriers, or the polarization of the incident field. In
some situation, one may need weaker conditions. For instance, one
may obtain the emission spectra with even harmonics only by using
the configuration of triple-dot (loop structure) with degenerate
levels for arbitrary inter-dot optical coupling. (It is a
consequence of the symmetry generated by $Q=\Omega\cdot\theta$ with
$\Omega$: $c_i \rightarrow -c_i$). Another issue is the preparation
of the appropriate initial state. Some initial state, for example,
that with electron staying in the lowest energy state, can be
obtained easily. One may use some special designed laser pulse to
prepare other types of initial states \cite{9Bavli1993PRA}. As also
seen in section II, small derivation of the initial conditions may
not change the picture. For instance, the main features of emission
spectra remain unchanged if the derivation of occupation probability
for the initial states is less than 5$\%$.


\section{\label{Sec:Con} Summary}
Based on the new theorem (on symmetry in the time
domain and energy spectrum domain) we proved, we provide  methods to
tailor photon emission patterns in driven nanostructures by tuning
the symmetry of the  system. Apart from the emission spectra with
only odd harmonics, or both odd and even harmonics, we are able to
obtain emission spectra with just even harmonics. In suitable
condition, the photon emission can even be fully quenched. Our
methods for tailoring emission spectra apply for general coupled
nanostructures.

\section{\label{Sec:Ack} Acknowledgments}
This work was partially supported by the
National Science Foundation of China under Grants No.10874020, No.
10774016, and by the National Basic Research Program of China (973
Program) under Grants No. 2011CB922204. We thank Dr. N. Yang for
helpful discussions.



\begin{thebibliography}{}
\bibitem{1Ahn2007PRL}
K. J. Ahn, F. Milde, and A. Knorr, Phys. Rev. Lett. \textbf{98},
027401 (2007).
\bibitem{2Chassagneux2009Nature}
Y. Chassagneux, Nature (London) \textbf{457}, 174 (2009).
\bibitem{3zhang2009PRB}
S.Q. Duan, W. Zhang, Y. Xie, W.D. Chu,and X.G. Zhao, Phys. Rev. B
\textbf{80}, 1 (2009).
\bibitem{agdot04}I. Baldea, Phys. Rev. B
 \textbf{69}, 245311 (2004).
\bibitem{terzis} A. F. Terzis, and E. Paspalakis, Phys. Rev. B
\textbf{80}, 035307 (2009); J. Appl. Phys. \textbf{97}, 023523
(2005).
\bibitem{moiseyev2001} V. Averbukh, O. E. Alon , and N. Moiseyev, Phys. Rev. A \textbf{64}, 033411
(2001).
\bibitem{4Ferrante2005PLA}
G. Ferrante, M. Zarcone, and S. A. Uryupin, Phys.Lett. A
\textbf{335}, 424 (2005).
\bibitem{symmetry}H. P. Breuer, K. Dietz, and M. Holthaus, Z. Phys. D \textbf{8}, 349
(1988); A. Peres, Phys. Rev. Lett. \textbf{67}, 158 (1991); F.
Grossmann, T. Dittrich, P. Jung, and P. Hanggi, Phys. Rev. Lett.
\textbf{67}, 516 (1991).
\bibitem{select} O. E. Alon, V. Averbukh, and N. Moiseyev, Phys. Rev. Lett. \textbf{80}, 3743 (1998);
Phys. Rev. Lett. \textbf{85}, 5218 (2000); I. Baldea, A. K. Gupta,
L. S. Cederbaum, and N. Moiseyev, Phys. Rev. B \textbf{69}, 245311
(2004); N. Ben-Tal, N. Moiseyev, and A. Beswick, J. Phys. B
\textbf{26}, 3017(1993); H. M. Nilsen, L. B. Madsen, and J. P.
Hansen, J. Phys. B \textbf{35}, L403 (2002).
\bibitem{5Xie2002PRA}
M. Xie, Phys. Res. A \textbf{483}, 527 (2002).
\bibitem{6Ferrante2004PLA}
G. Ferrante, M. Zarcone, and S. A. Uryupin, Phys.  Lett. A
\textbf{328}, 481 (2004).
\bibitem{9Bavli1993PRA}
R. Bavli and H. Metiu, Phys. Rev. A \textbf{47}, 3299 (1993).
\bibitem{10Thomas2001PRL}
T. Kreibich et al., Phys. Rev. Lett. \textbf{87}, 103901 (2001).
\bibitem{8Pietro2007MO}
P. P. Corso et al., J. Modern Opt. \textbf{54}, 1387 (2007).
\bibitem{11Wu2008OLA}
J. Wu, H. Qi and H. Zeng, Opt. lett. A \textbf{33}, 2050 (2008).
\bibitem{12Zhou2008PRA}
Z. Zhou and J. Yuan, Phys.Rev.A \textbf{77}, 063411 (2008).
\bibitem{13Heslar2007IJQC}
J. Heslar et al., Int. J. Quantum Chem. \textbf{107},3159 (2007).
\bibitem{14Le2007PRA}
V. H. Le et al., Phys.Rev.A \textbf{76}, 013414 (2007).
\bibitem{15F2006PRL}
F. Quere et al., Phys.Rev.lett. \textbf{96}, 125004 (2006).
\bibitem{16McPherson1987OSAB}
A. McPherson et al., J. Opt. Soc. Am. B \textbf{4}, 1753 (1987).
\bibitem{17Huillier1991PhysB}
A. L. Huillier, K. J. Schafer, and K. C. Kulander, J. Phys. B
\textbf{24}, 3315 (1991).
\bibitem{18Protopapas1997RPP}
M. Protopapas, C. H. Keitel, and P. L. Knight, Rep. Prog. Phys.
\textbf{60}, 389 (1997).
\bibitem{19Salieres1999AAMOP}
P. Sali¨¨res et al., Adv. At. Mol. Opt. Phys. \textbf{41}, 83
(1999).
\bibitem{20Chang1997PRL}
Z. Chang et al., Phys. Rev. Lett. \textbf{79}, 2967(1997).
\bibitem{21Schnurer1998PRL}
M. Schn¨¹rer et al., Phys. Rev. Lett. \textbf{80}, 3236 (1998).
\bibitem{22Salieres1998PRL}
P. Sali¨¨res et al., Phys. Rev. Lett. \textbf{81}, 5544 (1998).
\bibitem{corso98} P. P. Corso, L. L. Cascio, and F. Persico, Phys.Rev.A \textbf{58},1549 (1998).
\bibitem{23Narducci1990PRA}
L.M.Narducci et al., Phys.Rev.A \textbf{42}, 1630 (1990).
\bibitem{initial}M. L. Pons, R. Taieb, and A. Maquet, Phys. Rev. A \textbf{54}, 3634 (1996);
M. Frasca, Phys. Rev. A \textbf{60}, 573 (1999); V. Delgado and J.
M. G. Llorente, J. Phys. B \textbf{33}, 5403 (2000).
\bibitem{zhao96}
H. Wang and X.G. Zhao, J. Phys.: Cond. Matt. \textbf{8}, L285
(1996).


\end{thebibliography}
\end{document}